# High Energy Storage Efficiency and Large Electrocaloric Effect in Lead-Free BaTi$_{0.89}$Sn$_{0.11}$O$_3$ Ceramic

Soukaina Merselmiz[1], Zouhair Hanani[1,2], Daoud Mezzane[1,*], Matjaz Spreitzer[3], Andraž Bradeško[3], David Fabijan[3], Damjan Vengust[3], Lahoucine Hajji[1], Zahra Abkhar[1], Anna Razumnaya[4,5], Brigita Rožič[3], Igor A. Luk'yanchuk[4,6], and Zdravko Kutnjak[3]

[1] IMED-Lab, Cadi Ayyad University, Marrakesh, 40000, Morocco

[2] ICMCB, University of Bordeaux, Pessac, 33600, France

[3] Jozef Stefan Institute, Ljubljana, 1000, Slovenia

[4] LPMC, University of Picardy Jules Verne, Amiens, 80039, France

[5] Physics Faculty, Southern Federal University, Rostov-on-Don, 344090, Russia

[6] Department of Cybersecurity and Computer Engineering, Faculty of Automation and Information Technologies, Kyiv National University of Construction and Architecture, Kyiv, Ukraine

[*] Corresponding author: daoudmezzane@gmail.com

## Abstract

Lead-free BaTi$_{0.89}$Sn$_{0.11}$O$_3$ (BTSn) ceramic was elaborated via a solid-state reaction method and its dielectric, ferroelectric, energy storage, electromechanical as well as electrocaloric properties were investigated at 25 kV/cm. Pure perovskite structure was confirmed by X-ray diffraction analysis. The maximum of the dielectric constant was found to be 17390 at 41 °C. The enhanced total energy density, the recovered energy density, and the energy storage efficiency of 92.7 mJ/cm$^3$, 84.4 mJ/cm$^3$, and 91.04%, respectively, were observed at 60 °C, whereas the highest storage efficiency of 95.87 % was obtained at 100 °C. At room temperature, the electromechanical strain and the large-signal piezoelectric coefficient reached a maximum of 0.07 % and 280 pm/V. The large electrocaloric effect of 0.71 K and the electrocaloric responsivity of 0.28 × 10$^{-6}$ K.mm/kV at 49 °C under 25 kV/cm were indirectly calculated via Maxwell relation from the ferroelectric polarization $P$ ($T$, $E$) that was determined from the $P$-$E$ hysteresis loops. By exploiting the Landau-Ginzburg-Devonshire (LGD) phenomenological theory, the electrocaloric response was estimated to be of 0.61 K at 50 °C under 25 kV/cm. We conclude that BTSn lead-free ceramic is a promising candidate for potential applications in high-efficiency energy storage devices and solid-state refrigeration technology.

**Keywords**: lead-free ceramic, energy storage, high-efficiency, piezoelectric, electrocaloric effect.

# 1. Introduction

Nowadays, there is a growing demand for small-scale refrigeration and energy-conversion technologies [1–3]. The recent researches are focusing on the design of new ferroelectric materials with high piezoelectric, electrocaloric, and energy-storage characteristics near room temperature. The ceramic-based dielectric capacitors can offer high energy density, fast charge/discharge speed, and long-cycle life in comparison with batteries and electromechanical capacitors [4–6]. At the same time, a piezoelectric effect of these materials allows the electromechanical coupling between the electrical field and mechanical strain [7]. Two types of the piezoelectric effects are explored, the direct piezoelectric effect, and the converse piezoelectric effect. In the former case, the voltage is generated after applying a mechanical strain on the piezoelectric material. In the latter case, the strain of the material is generated under the application of an external electric field. The electrocaloric (EC) effect is emerging recently as a not yet fully explored phenomenon in ferroelectric materials that can lead to the development of new cooling technologies [8,9]. Currently, solid-state cooling devices are in progress to replace the vapor compression cycle (VCC) based devices, because they have more eco-friendly potential and could reach higher efficiency [8]. The electrocaloric effect of the dielectric material consists of the producing of an isothermal entropy change, $\Delta S$, upon applying or removing an external electric field. Consequently, an adiabatic temperature change, $\Delta T$, is induced [10]. Importantly, the ferroelectrics are considered as potential candidates with significant EC effect due to their large pyroelectric coefficient [11]. Moreover, the electrocaloric effect could be investigated by direct EC measurements by using a plethora of investigation techniques that enable measurement of the material temperature change [9,12–14], the indirect EC effect calculation through Maxwell relation [13–15] and theoretical investigations based on the phenomenological approach via Landau-Ginzburg-Devonshire (LGD) theory [16].

Over half a century, lead-based ferroelectric perovskite materials Pb(Zr,Ti)O$_3$ (PZT) have dominated applications since they have been commercially used in numerous electronic devices, such as resonators, ultrasonic generators, actuators, due to their excellent electromechanical, ferroelectric and energy storage properties [3,17,18]. At the same time, the lead-containing ferroelectric materials are suffering from the health safety issue related to the lead-caused high-toxicity [17,19,20], despite their remarkable properties. Consequently, there is a current upsurge in research on the development of lead-free ferroelectric materials with comparable properties to replace lead-based materials [21]. One of generic lead-free material is a barium titanate BaTiO$_3$ (BTO) ceramic which is considered to be one of the environmentally friendly materials that can be used in the development of novel lead-free materials with high piezoelectric, electrocaloric and energy storage properties [21–23]. However, the BTO ceramics display a low dielectric constant, sharp ferroelectric-paraelectric phase transition at higher Curie temperature, $T_c$ = 120 °C, and low piezoelectric coefficient, $d_{33}$ = 190 pC/N [24]. To improve its properties, the BTO perovskite structure can be tailored by site engineering [15,25]. For example, BaTi$_{(1-x)}$Sn$_x$O$_3$ (denoted as BTSn) system was designed by the insertion of Sn$^{4+}$ into B-site of BaTiO$_3$ crystal lattice to replace Ti$^{4+}$. It should be also noted that such system could possess an extremely high dielectric constant. Yao et al. [26] elaborated BaTiO$_3$-xBaSnO$_3$ system by solid-state reaction route, and found that BT-11BS ceramics presented excellent properties; high dielectric constant, $\varepsilon_r$ = 75000, at $T_c$ = 42 °C and large piezoelectric coefficient, $d_{33}$ = 697 pC/N, due to the coexistence of the four phases, cubic, orthorhombic, tetragonal, and rhombohedral (*C, T, O, R*) at a quasi-quadruple point, $Q_p$, existing near the room temperature. Besides, Sanlialp et al. [27] prepared BTSn (x = 0.11) and reported a significant electrocaloric effect with high temperature change of 0.63 K at 20 kV/cm and at 44 °C utilizing direct measurement via differential scanning calorimeter. Gao et al. [28] evaluated the dielectric properties and the energy storage in BTSn ceramic with x = 0.105 and found that the dielectric constant and the high energy storage density reached 54000 and 30 mJ/cm$^3$ at 10 kV/cm, respectively. It is worth to mention that the evaluation of the EC effect by indirect method following Maxwell relation is frequently employed in ferroelectric materials [13,27,28-30]. However, there is still some debate on the validity of the EC effect deduced from the Maxwell relation in relaxor ferroelectrics, because they are not in thermal or mechanical equilibrium [31]. Additional estimations of the EC effect in ferroelectric materials can be obtained by theoretical investigations exploiting the LGD phenomenological approach [16].

According to the literature, BT$_{(1-x)}$Sn$_x$O$_3$ ceramic with x = 0.11 exhibits a large dielectric constant and high energy storage properties under low electric field at $Q_p$ near the room temperature, owing to the vanishing of energy barrier around $Q_p$. This energy reduction facilitates polarization rotation and makes the material easily polarized [28,32]. In the current work, we aim to study the temperature-dependence of energy storage, electromechanical and electrocaloric properties in BTSn (x = 0.11) ceramic under moderate electric field of 25 kV/cm simultaneously. The electrocaloric effect was investigated via two methods: indirect approach following Maxwell relation and phenomenological approach through LGD theory. Comparing the obtained results with the literature data we conclude that the studied BTSn ceramic is a suitable material for applications in solid-state refrigeration technologies and energy storage devices.

## 2. Experimental

### 2.1. Synthesis of BTSn ceramic

Lead-free $BaTi_{0.89}Sn_{0.11}O_3$ (BTSn) ceramic was elaborated through the conventional solid-state reaction route. The stoichiometric amounts of $BaCO_3$, $TiO_2$ and $SnO_2$ were mixed with ethanol as a medium in an agate mortar for 2h, then dried and calcined at 1200 °C for 12h with a heating rate of 5 °C/min. The calcined powders were uniaxially pressed into pellets of diameter about 12 mm and thickness about 1 mm, using 5 wt % of polyvinyl alcohol (PVA) as a binder. Therefore, the pellets were first heated up at 800 °C for 2h to burn out the binder and then sintered at 1350 °C for 7h.

## 2.2. Characterizations

The room-temperature crystalline structure of BTSn sintered ceramic was determined by the X-ray diffraction (XRD, Panalytical X-Pert Pro) under a step angle of 0.02° in the $2\theta$ range from 10° to 80° using Cu-$K_\alpha$ radiation ($\lambda \sim 1.5406$ Å). The lattice parameters of the sample were refined by using FullProf software. The surface morphology of the sintered ceramic was examined by using the Scanning Electron Microscopy (SEM, Tescan VEGA3). The grain size distributions of the sample were determined by using ImageJ software. The density of the sintered ceramic was evaluated by the Archimedes method using deionized water as a medium. The dielectric properties of BTSn ceramic with gold paste electrodes were measured by using a precision LCR Meter (Agilent, 4284A) in the frequency range of 1 kHz to 1 MHz. The polarization–electric field ($P$–$E$) and strain–electric field ($S$–$E$) hysteresis loops were simultaneously measured by using an AixACCT TF 2000 Analyzer with a SIOS Meßtechnik GmbH laser interferometer and a TREK model 609E-6 high-voltage amplifier. The hysteresis loops were measured by using an excitation sinusoidal signal with a frequency of 10 Hz in the temperature range of 30-130 °C with a 5 °C step upon the heating cycle. The temperature-dependence of the electrocaloric temperature change ($\Delta T$) and responsivity ($\zeta$) were calculated in the indirect method from the recorded $P$–$E$ hysteresis loops at 10 Hz.

# 3. Results and discussions

## 3.1. Structural properties

The room temperature XRD pattern of BTSn ceramic sintered at 1350 °C/7h is plotted in Fig. 1a. The sample displays a pure perovskite phase without any secondary phase. Also, by using Fullprof Suite software, the Rietveld refinement was carried out showing a tetragonal phase with *P4mm* space group which is branded by the splitting of both (002)/(200) peaks around $2\theta \approx 45°$ and (202)/(220) peaks at around $2\theta \approx 65°−66°$ as shown in Fig. 1b (insets). Besides, $Y_{obs}$ and $Y_{calc}$ represent the observed and calculated data, respectively. It is worth to mention that due to sintering at the higher temperature, no other secondary phases were detected which confirm the complete solid solubility of $Sn^{4+}$ at Ti-site. The lattice parameters, the atomic positions, and the quality-of-fit measure ($\chi^2$) obtained by the refinement are gathered in Table 1.

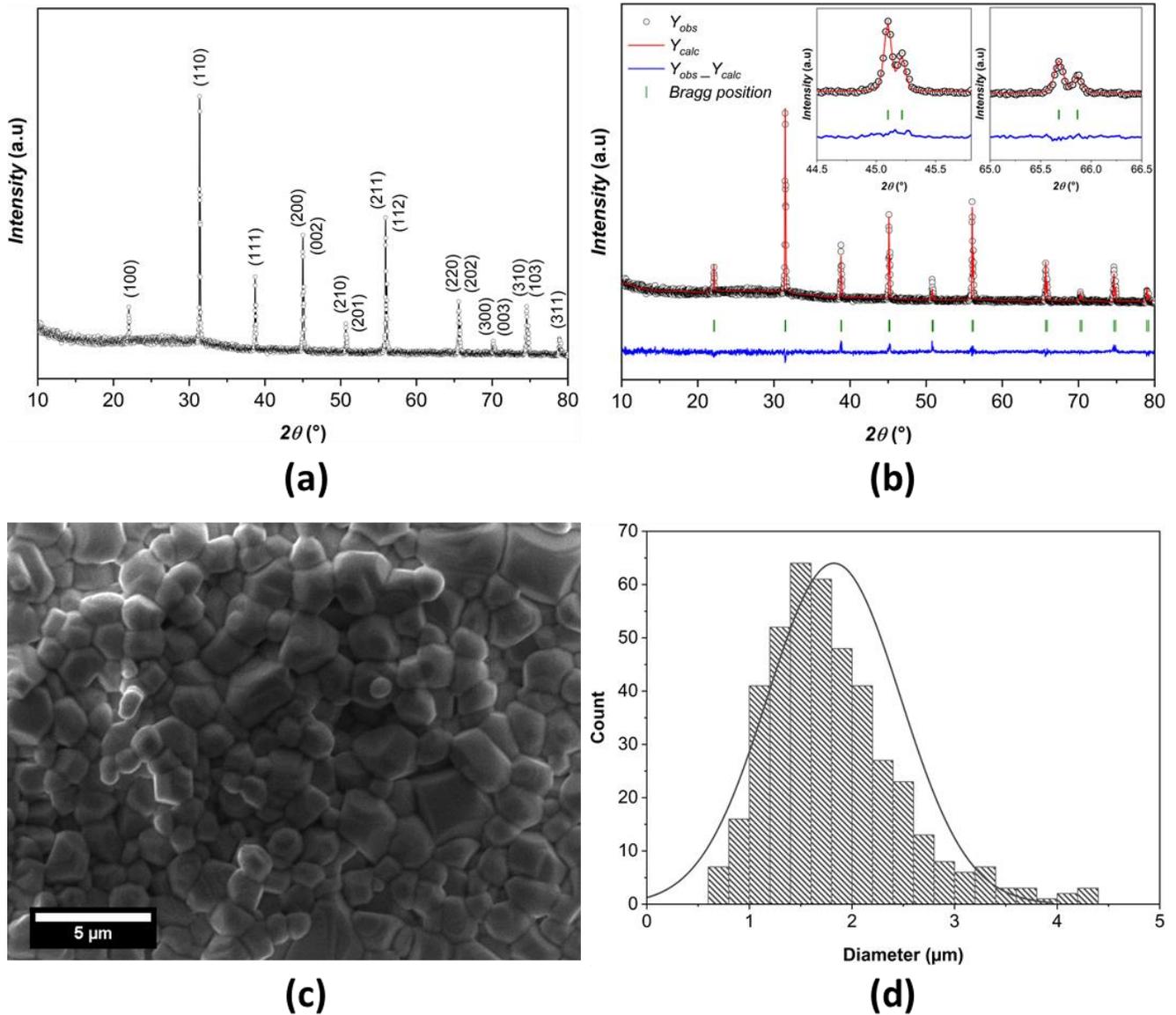

**Fig. 1.** (a) Room-temperature XRD pattern (b) Rietveld refinement, (c) SEM micrograph, and (d) grain size distribution of BTSn ceramic sintered at 1350 °C/7h.

Fig. 1 (c-d) illustrates the SEM micrograph and grain distribution of BTSn ceramic sintered at 1350 °C/7h. The sample exhibits a well-connected and dense microstructure. It is worth to mention that BTSn sample presents a few facetted grains with small and coarse grains. Therefore, the Gaussian distribution of grain size characterized by a mean grain size of $1.82 \pm 0.64$ µm was observed, and the material density was found to be 5.50 g/cm³.

**Table 1.** Refined structural parameters listed for BTSn ceramic at room temperature.

| Lattice parameter (Å) | Angle (°) | Volume (Å³) | Space group | $c/a$ | Atomic position (x, y, z) | | | | $\chi^2$ |
|---|---|---|---|---|---|---|---|---|---|
| $a = 4.01696$ | $\alpha = \beta = \gamma$ | 64.82 | P4mm | 1.00017 | Ba | 0.0000 | 0.0000 | 0.0221 | 1.721 |
| $b = 4.01696$ | $= 90$ | | | | Ti/Sn | 0.5000 | 0.5000 | 0.4810 | |
| $c = 4.01768$ | | | | | O1 | 0.5000 | 0.5000 | 0.7226 | |
| | | | | | O2 | 0.5000 | 0.5000 | 0.0483 | |

### 3.2. Dielectric properties

Fig. 2. (a) displays room-temperature frequency dependence of the dielectric constant ($\varepsilon_r$) and the dielectric loss ($tan\ \delta$) in the selected frequency window from 1 kHz to 1 MHz in BTSn ceramic. It was observed that the dielectric properties of BTSn are strongly dependent on the frequency. $\varepsilon_r$ decreases continuously with increasing the frequency which is attributed to the attenuation of space charge polarization effect (interfacial polarization). However, $tan\ \delta$ remains small at low frequencies, then, rapidly increases at higher frequencies ($>10^5$ Hz). At low frequencies, BTSn ceramic exhibits well defined conducting grains separated by insulating grain boundaries, which causes accumulation of charges under the influence of the electric field and thus increasing the interfacial polarization resulting in a high dielectric constant. Nevertheless, at higher frequencies, the dipoles cannot align themselves in the direction of the electric field resulting in a reduced interfacial polarization and correspondingly low dielectric constant and high dielectric loss at high frequencies [33]. At 1 kHz, a high dielectric constant of 11025 was found, which is higher than dielectric constant of some other lead-free ceramics such as BCZT [32,34,35].

Fig.2 (b) illustrates the temperature-dependence of $\varepsilon_r$ and $tan\ \delta$ at various frequencies in BTSn sample from -40 to 200 °C. The sample exhibits only one dielectric anomaly associated with the direct cubic-rhombohedral ($C$-$R$) phase transition. This dielectric transition is regularly named as pinched transition temperature in literature, and was related to the coexistence of four phases (cubic-tetragonal-orthorhombic-rhombohedral or $C$-$T$-$O$-$R$) at a quadruple point $Q_p$ [26,36]. It is interesting to note that at 1 kHz, the BTSn sample shows a high value of the maximum dielectric constant ($\varepsilon_m$) equal to 17390 around 41 °C, with significant dielectric dispersion, indicating the feature in relaxor ferroelectrics.

Besides, the temperature-dependence of the dielectric loss exhibits a step-like behavior, highlighting the ferroelectric-paraelectric transition (FE-PE) near the room temperature as shown in Fig. 2b. At the transition peak and 1 kHz, *tanδ* was found to be 0.055. These results reveal that Sn doping BaTiO$_3$ [24] decreases $T_c$ strongly, but strengthens the dielectric properties [36,37]. Furthermore, based on the LGD theory analysis, the dielectric constant enhancement at $Q_p$ originates from vanishing energy barriers for polarization rotation and extension [26,28]. The obtained dielectric results are similar to that of Jin et al. [38] where BTSn11 ceramic was synthesized by solid-state method.

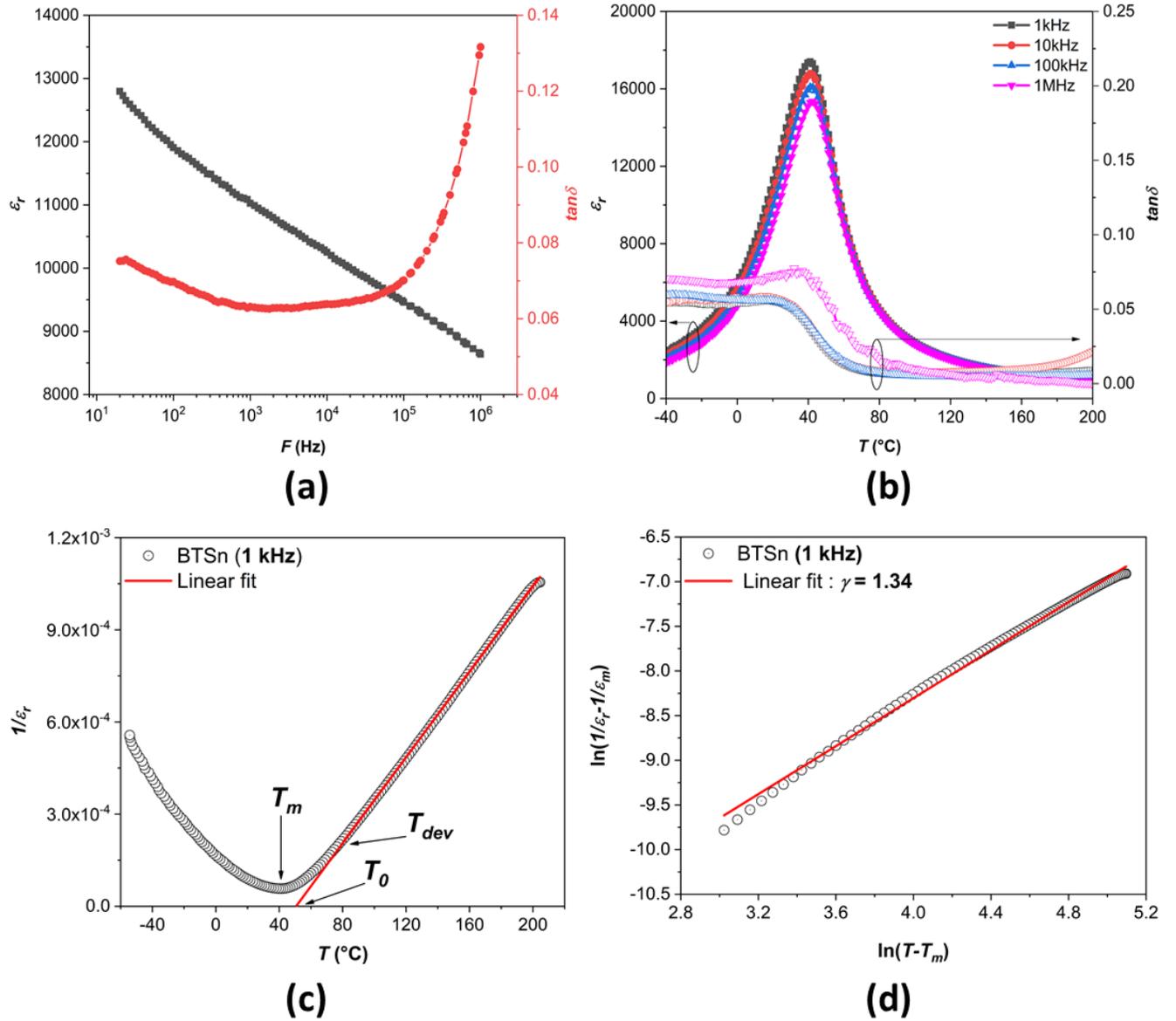

**Fig. 2.** (a) Room-temperature frequency-dependence of the dielectric constant and dielectric loss, (b) temperature-dependence of the dielectric constant and dielectric loss at different frequencies, (c) plot

showing the Curie-Weiss relation and (d) fit to modified Curie-Weiss law with slope $\gamma$=1.34 of BTSn ceramics ceramic sintered at 1350 °C/7h.

It is worth noting that the peak temperatures are frequency-dependent and shifted toward high temperature with increasing frequency (reminiscent of relaxor behavior) [39,40]. To thoroughly investigate this phase transition, the inverse dielectric constant as a function of temperature was depicted and fitted at 1 kHz in Fig. 2c by using the Curie–Weiss law:

$$\frac{1}{\varepsilon_r} = \frac{T - T_0}{C} (T > T_0), \qquad (1)$$

where $\varepsilon_r$ is the real part of the dielectric constant, $T_0$ is the Curie Weiss temperature, and $C$ is the Curie-Weiss constant.

The fitting results found for BTSn ceramic by using Eq. (1) are summarized in Table 2. It is worth to mention that the dielectric constant ($\varepsilon_r$) deviates from the Curie–Weiss law above the $T_c$. These deviations $\Delta T_m$ was determined by using Eq. (2) [41]:

$$\Delta T_m(K) = T_{dev} - T_m, \qquad (2)$$

where, $T_{dev}$ designates the temperature below which $\varepsilon_r$ starts to deviate from the Curie–Weiss law, and $T_m$ indicates the temperature at which $\varepsilon_r$ value attained the maximum. The calculated $\Delta T_m$ of BTSn ceramic is equal to 39 °C.

The Curie constant value in order of $10^5$ K indicates that BTSn sample is of the displacive-type ferroelectric such as BaTiO$_3$ (1.7×10$^5$ K) [42] that exhibits a displacive ferroelectric phase transition. Eq. (3) was employed to extract the phase transition diffuseness [43]:

$$\frac{1}{\varepsilon_r} - \frac{1}{\varepsilon_m} = \frac{(T - T_0)^\gamma}{C} (1 < \gamma < 2), \qquad (3)$$

where $\varepsilon_r$, $\varepsilon_m$ are the real part of the dielectric constant and its maximum value, respectively, and $\gamma$ (degree of transition diffuseness) and $C$ are constant. It is worth to mention that for a normal ferroelectric (non-diffuse, Curie-Weiss law) $\gamma = 1$, however for an ideal relaxor ferroelectric (diffuse, Smolenskii–Isupov law) $\gamma = 2$ [44]. Fig. 2d plots the linear relationship between $ln$ ($1/\varepsilon_r$ -$1/\varepsilon_m$) and $ln$ ($T - T_m$). By curve fitting employing Eq. (3), the value found for $\gamma$ at 1 kHz in BTSn ceramic is 1.34.

**Table 2.** Relaxor properties at 1 kHz of BTSn ceramic sintered at 1350 °C/7h.

| | $\varepsilon_m$ | $\tan \delta$ | *Grain size* (μm) | $T_0$ (°C) | $C \times 10^5$ (K) | $T_m$ (°C) | $T_{dev}$ (°C) | $\Delta T_m$ (°C) | $\gamma$ |
|---|---|---|---|---|---|---|---|---|---|
| **BTSn** | 17390 | 0.055 | $1.82 \pm 0.64$ | 51 | 1.44 | 41 | 80 | 39 | 1.34 |

### 3.3. Energy storage properties

To highlight the ferroelectric properties of BTSn sample, the bipolar *P-E* hysteresis loops obtained under 25 kV/cm at the driving frequency of 10 Hz are shown in Fig. 3a. At room temperature, a typical *P-E* hysteresis loop was observed, and the maximal polarization ($P_{max}$), the remnant polarization ($P_r$), the charge storage density ($Q_c = P_{max} - P_r$) and the coercive field ($E_c$) were found to be 14.16 μC/cm², 3.24 μC/cm², 10.92 μC/cm², and 1.2 kV/cm, respectively. With increasing temperature, the *P-E* loops turn out to be gradually slimmer, accompanied by a decrease of $P_{max}$, $P_r$, $Q_c$ as well as $E_c$ (Fig. 3b), owing to the disappearance of ferroelectric domains. Also, the linear P-E trend and absence of the hysteresis loops above the Curie temperature, confirm the characteristic of pure paraelectric phase. It is worthy to mention that in the ferroelectric capacitor, the charge storage density ($Q_c$) is determined from the *P-E* hysteresis loops at zero fields and high $Q_c$ value is needed for high energy density [45].

To further investigate the energy storage properties in BTSn ceramic, the *P-E* hysteresis loops as a function of temperature were recorded. For nonlinear dielectrics such as ferroelectric relaxors, which possess some energy dissipation, the total energy density ($W_{tot}$) and recoverable energy density ($W_{rec}$) could be estimated through the equations (4) and (5). $W_{tot}$ can be calculated by adding the area gathering $W_{rec}$ and energy loss density ($W_{loss}$) as schematically presented in Fig. 3c. Moreover, $W_{rec}$ could be determined by integrating the area between the polarization axis and the upper branch curve of the unipolar *P-E* hysteresis loop (blue area), whereas, $W_{loss}$ produced by the domain reorientation is determined by integrating the *P-E* loop area (red area) in Fig. 3c [15,46,47].

$$W_{tot} = \int_0^{P_{max}} E \, dP, \qquad (4)$$

$$W_{rec} = \int_{P_r}^{P_{max}} E \, dP, \qquad (5)$$

To reach high energy storage in ferroelectric materials, high dielectric constant, maximal polarization, small remnant polarization, and high breakdown strength are required [4,46]. Besides, the energy storage efficiency ($\eta$) is an important parameter for evaluating the energy storage performances which is defined in Eq. (6). During the depolarization process, a part of the stored energy ($W_{tot}$) would be dissipated and presented as energy loss density. After applying a high electric field to get high energy density, an increase in the electronic/ionic conductivity of the sample results in an increase of dielectric loss (*tan $\delta$*), finally reducing the energy storage efficiency. Consequently, the energy dissipated by the hysteresis loss ($U_H$) (Eq. (7)) will lead to temperature rise of the capacitor, reducing its thermal stability and lifetime [46].

$$\eta \ (\%) = \frac{W_{rec}}{W_{tot}} \times 100 = \frac{W_{rec}}{W_{rec} + W_{loss}} \times 100, \qquad (6)$$

$$U_H = \pi \varepsilon_0 \varepsilon_r E^2 tan\delta, \qquad (7)$$

Fig. 3d depicts the temperature-dependence of the energy storage performances of BTSn ceramic under 25 kV/cm. At room temperature $W_{tot}$, $W_{rec}$ and $\eta$ were found to be 85.1, 72.4 mJ/cm$^3$, and 85.07 %, respectively. However, with increasing the temperature, $W_{tot}$ and $W_{rec}$ gradually increase to reach a maximum of 92.7 and 84.4 mJ/cm$^3$, respectively, around 60 °C, then diminish gradually. Accordingly, at 60 °C, high energy storage efficiency of 91.04 % was reached in BTSn ceramic. Furthermore, at 100 °C, the maximum of $\eta$ was found to be 95.87 % and associated with $W_{rec}$ equal to 65.1 mJ/cm$^3$. It was observed that $W_{tot}$, $W_{rec}$ curves present the same trend in comparison to the energy storage efficiency ($\eta$) while increasing the temperature. This behavior could be associated to the hysteresis loss, because with increasing temperature, the *P-E* hysteresis loops become slimmer, and accordingly $W_{rec}$ increases and $W_{loss}$ decreases. Consequently, the ratio of the energy storage efficiency increases to reach a maximum at 100 °C where the *P-E* hysteresis loops are linear (see Fig. 3a).

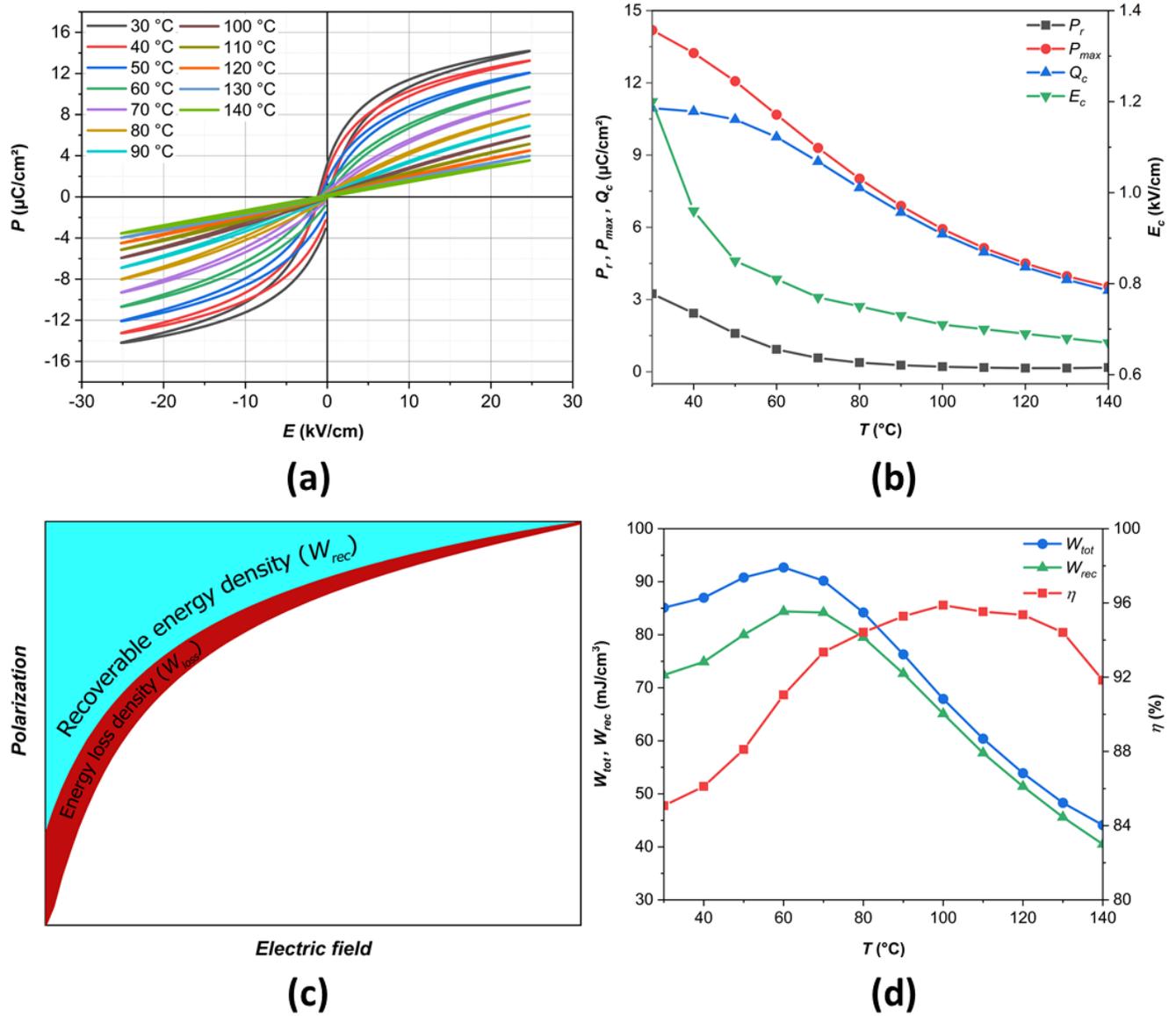

**Fig. 3.** Temperature-dependence of (a) $P$-$E$ loops, (b) $P_r$, $P_{max}$, $Q_c$, and $E_c$.(c) A schematic depiction of the relevant energy storage energies determined via $P$-$E$ hysteresis loops (d) The relevant energy storage parameters of BTSn ceramic as a function of temperature.

Table 3 summarizes the comparison of our results with energy storage properties of several lead-free ceramics. It is worth to mention that applying a large electric field induces higher polarization, thus, the energy storage capabilities increase. Furthermore, the enhanced recoverable energy density is attributed to the large $P_{max}$ and high dielectric breakdown strength (*DBS*). For instance, Gao et al. [28] prepared BTSn10.5 ceramic by solid-state method, and observed an energy storage density of 31 mJ/cm³ under a low electric field of 10 kV/cm. In addition, under a low electric field of 6.5 kV/cm, Hanani et al. [15] reported a high energy storage efficiency ($\eta$) of 80 % at 120 °C, associated with $W_{tot}$ and $W_{rec}$ equal to 16.5 and 14 mJ/cm³, respectively. Nevertheless, under a high electric field of 170 kV/cm, Puli et al.

[48] stated a high energy storage properties ($W_{tot}$ = 680 mJ/cm³, $W_{rec}$ = 495 mJ/cm³ and $\eta$ = 72.8 %) for BZT-x BCT ceramics. However, the energy storage efficiency of the presented literature data is lower in comparison to our findings. Consequently, BTSn ceramic could be suitable for the design of energy storage systems with high-efficiency.

**Table 3.** Comparison of the energy storage properties of BTSn ceramic with other lead-free ceramics reported in literature.

| Ceramic | $W_{tot}$ (mJ/cm³) | $W_{rec}$ (mJ/cm³) | $\eta$ (%) | $E$ (kV/cm) | $T$ (°C) | Ref. |
|---|---|---|---|---|---|---|
| BTSn 11 | 67.9 | 65.1 | 95.87 | 25 | 100 | This work |
| BTSn 11 | 92.7 | 84.4 | 91.04 | 25 | 60 | This work |
| BTSn 11 | 85.1 | 72.4 | 85.07 | 25 | 30 | This work |
| BTSn 10.5 | 31 | - | - | 10 | 40 | [28] |
| BaTiO$_3$ | - | 450 | 28.23 | 110 | - | [49] |
| BZT | 302 | 218 | 72 | 50 | - | [50] |
| Ba$_{0.95}$Ca$_{0.05}$Zr$_{0.3}$Ti$_{0.7}$O$_3$ | 590 | 429 | 72.8 | 160 | - | [51] |
| Ba$_{0.955}$Ca$_{0.045}$Zr$_{0.17}$Ti$_{0.83}$O$_3$ | 680 | 495 | 72.8 | 170 | - | [48] |
| Ba$_{0.85}$Ca$_{0.15}$Zr$_{0.10}$Ti$_{0.90}$O$_3$ | 16.5 | 14 | 80 | 6.5 | 120 | [15] |

### 3.4. Piezoelectric properties

*S-E* curves were recorded under 25 kV/cm at the driving frequency of 10 Hz. Fig. 4a depicts the field-induced bipolar strain *S-E* hysteresis curves for BTSn ceramic recorded from 30 °C to 100 °C. At 30 °C, typical butterfly-shaped *S-E* curves were observed, showing a purely positive strain, and an enhanced average strain $S_{ave}$ of 0.07 % was obtained. This butterfly-loop behavior is due to the electromechanical strain of the lattice related to the switching and movement of domain walls by the electric field [52,53].

It is worth to mention that at room temperature, the *S-E* curves exhibit an asymmetric nature for the positive and negative electric field, which may be associated with the internal bias field arising from defects [54,55]. As the temperature increases, the *S-E* curves become slimmer and sprout-shaped, indicating a pure paraelectric phase. The large-signal piezoelectric coefficient also denoted as normalized strain is calculated using the following equation,

$$d_{33}^* = \frac{S_{max}}{E_{max}}. \qquad (8)$$

Here $S_{max}$ is the maximum strain measured at the maximum electric field $E_{max}$. Fig. 4b shows the thermal evolution of $d_{33,ave}^*$ (the average value of the $d_{33}^*$ at the positive and negative electric field). At 30 °C, $d_{33,ave}^*$ reaches a maximum of 280 pm/V, and gradually decreases with increasing temperature due to the transition from ferroelectric to paraelectric phase.

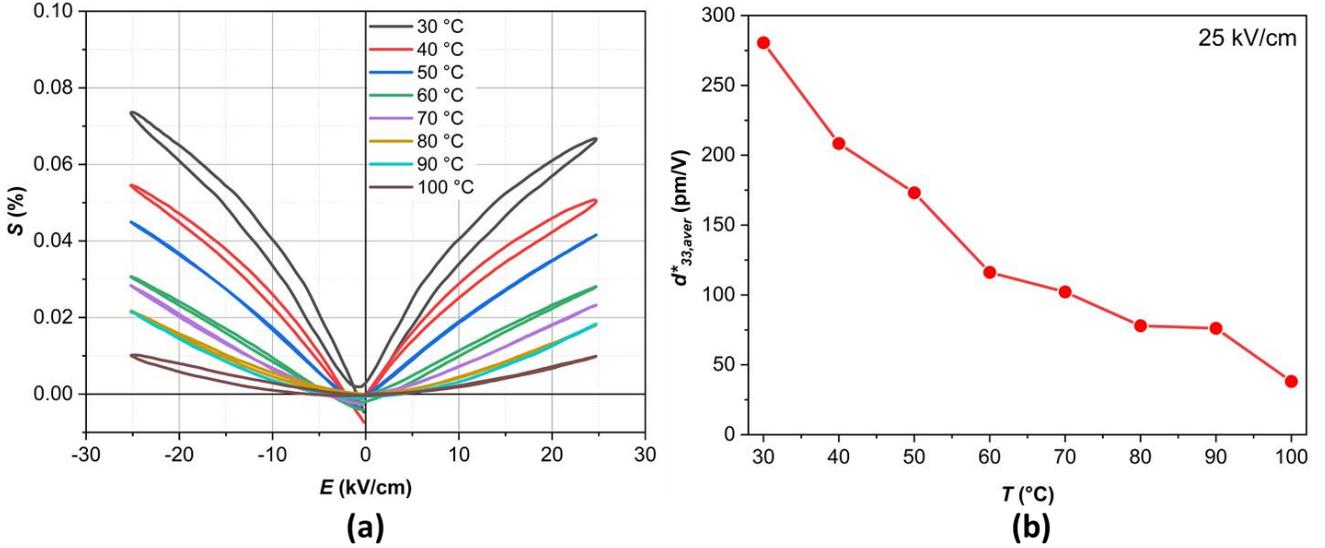

**Fig. 4.** Temperature-dependence of (a) the field-induced bipolar strain hysteresis loops (*S-E*) and (b) the large signal piezoelectric coefficient in BTSn ceramic.

Table 4 lists the comparison of the electromechanical properties of BTSn and several lead-free ferroelectric ceramics at room temperature. BTSn ceramic exhibits an average strain $S_{ave}$ of 0.07 % according to an enhanced $d_{33,ave}^*$ of 280 pm/V under 25 kV/cm. Chaiyo et al. [56] stated that $Ba_{0.75}Ca_{0.25}TiO_3$ ceramic prepared by solid-state method, displays a high strain of 0.519 % with a maximum $d_{33}^*$ of 173 pm/V at 30 kV/cm. Moreover, they elaborated also $Ba_{0.80}Ca_{0.20}Zr_{0.05}Ti_{0.95}O_3$ ceramic which shows a high strain of 0.852 % associated with a comparable $d_{33}^*$ of 284 pm/V. At the same electric field, higher $d_{33}^*$ of 513 pm/V was found in BZT-50BCT ceramic prepared by the sol-gel method as reported by Praveen et al. [41]. Moreover, Jin et al. [38] synthesized BTSn11 by solid-state method and reported that under a high electric field of 40 kV/cm, a comparable strain of 0.083% and $d_{33}^*$ of 208 pm/V were found at room temperature. The differences in these results could be ascribed to the different elaboration methods, chemical compositions, site occupancy, the applied electric field, and the measurement conditions.

**Table 4.** Comparison of the electromechanical properties of BTSn ceramic and other lead-free ferroelectric ceramics at room temperature.

| Ceramic | $T$ | $S_{max}$ | $E_{max}$ | $d_{33}^*$ | Ref. |
|---------|-----|-----------|-----------|-----------|------|
| | (°C) | (%) | (kV/cm) | (pm/V) | |
| BTSn | 30 | 0.07 | 25 | 280 | This work |
| $Ba_{0.75}Ca_{0.25}TiO_3$ | 30 | 0.519 | 30 | 173 | [56] |
| $Ba_{0.80}Ca_{0.20}Zr_{0.05}Ti_{0.95}O_3$ | 30 | 0.852 | 30 | 284 | [56] |
| BZT-52BCT | 40 | 0.076 | 30 | 250 | [41] |
| BZT-50BCT | 40 | 0.157 | 30 | 513 | [41] |
| BTSn11 | 30 | 0.083 | 40 | 208 | [38] |
| $Ba_{0.85}Ca_{0.15}Zr_{0.08}Ti_{0.92}O_3$ | 25 | 0.185 | 40 | 463 | [57] |

## 3.5. Electrocaloric properties

To estimate the EC effect in lead-free BTSn ceramic, the indirect method using Maxwell relation and Landau-Ginzburg-Devonshire (LGD) phenomenological theory was employed, and the results were compared.

### 3.5.1 Indirect approach following Maxwell relation

The indirect method was used based on the measured ferroelectric order parameter $P$ $(T, E)$ determined from the $P$-$E$ hysteresis loops as a function of temperature. From these $P$-$E$ hysteresis curves (see Fig 3a), a seven-order polynomial fitting of upper branches was carried out at each applied electric field. The continually decreasing above $T_c$ thermal evolution of the polarization is plotted in Fig. 5a

The reversible electrocaloric temperature change ($\Delta T$) was determined through an indirect approach based on the Maxwell equation,

$$\Delta T = -\int_{E_1}^{E_2} \frac{T}{\rho C_p} \left(\frac{\partial P}{\partial T}\right)_E dE. \qquad (9)$$

Here $\rho$ and $C_p$ are the mass density and the specific heat of the material, respectively.

The dependencies of the temperature change, $\Delta T$ on the temperature $T$, calculated at different electric fields, are shown in Fig. 5b. All curves exhibit the maximums at the ferroelectric-paraelectric phase transition which are increasing with increasing of the applied electric field. The value of $\Delta T$ reaches a maximum of 0.71 K at $T$=52 °C and $E$=25 kV/cm, indicating a few degrees shift up of the corresponding $Q_p$. It is worth to mention that the EC responsivity $\zeta = \Delta T / \Delta E = 0.284$ K.mm/kV for BTSn ceramic at 25 kV/cm.

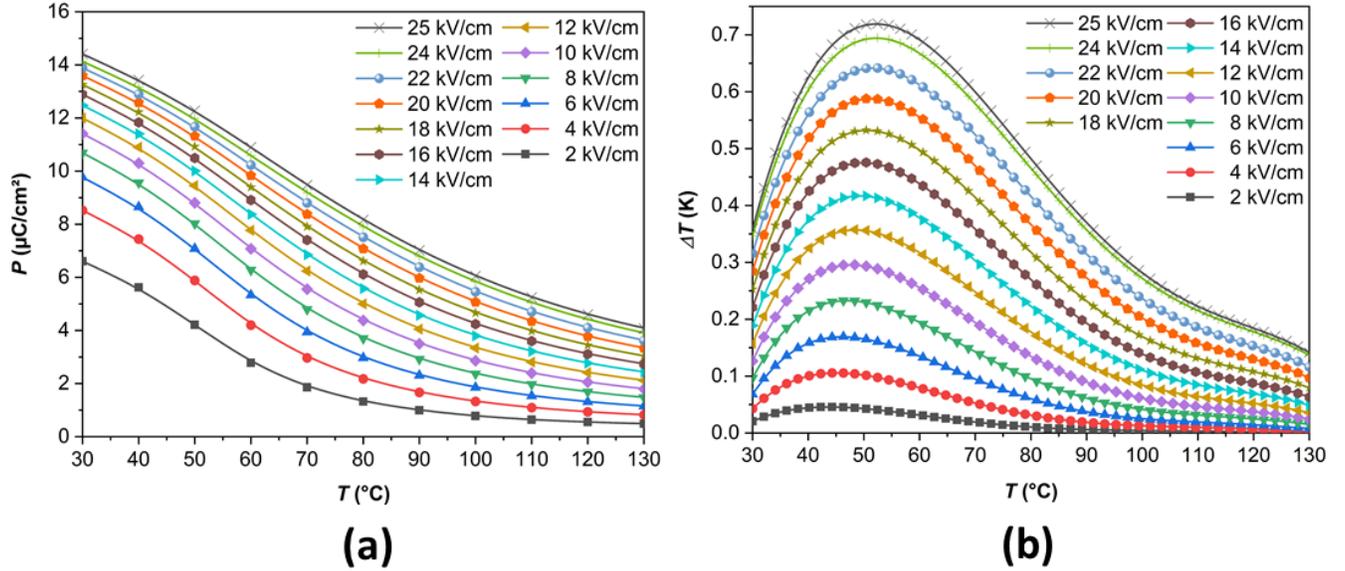

**Fig. 5.** Temperature-dependence of (a) polarization and (b) electrocaloric temperature change ($\Delta T$) determined at different applied electric fields in BTSn ceramics.

Table 5 compares the EC response ($\Delta T$ and $\zeta$) of the BTSn ceramic with previously published results obtained in lead-free ferroelectric materials. At $E$=25 kV/cm, the EC response of BTSn ceramic obtained by indirect approach ($\Delta T = 0.71$ K, $\zeta = 0.284$ K.mm/kV), are highest in comparison to those reported in Table 5. At the same applied electric field, Ba$_{0.90}$Ca$_{0.10}$Zr$_{0.05}$Ti$_{0.95}$O$_3$ ceramic attained the maximum EC response ($\Delta T = 0.465$ K, $\zeta = 0.186$ K.mm/kV) at 119 °C, as reported by Ben Abdessalem et al. [30]. Nevertheless, at 20 kV/cm, Luo et al. [58] found comparable $\Delta T$ and $\zeta$ values of 0.61 K and 0.31 K.mm/kV, respectively, at the a quasi-quadruple point $Q_p$ in BTSn ceramic with x = 0.105. Furthermore, Sanlialp et al. [27] studied the EC response by means of direct approach in BTSn ceramic with x = 0.11 ($\Delta T = 0.63$ K, $\zeta = 0.315$ K.mm/kV) and BTSn ceramic with x = 0.14 ($\Delta T = 0.38$ K, $\zeta = 0.19$ K.mm/kV), showing that while increasing Sn$^{4+}$ ions, the EC response decreases. The obtained value in BTSn with x = 0.11, found by Sanlialp et al. [27] via the direct method is similar to that obtained in our study, which indicates that the indirect approach via Maxwell equation (Eq. (9)) is reliable method to evaluate the EC response. Hanani et al. [29] elaborated rod-like ceramics by surfactant-assisted solvothermal processing and found $\Delta T$ and $\zeta$ of 0.492 K and 0.289 K.mm/kV, respectively, around 87 °C. The enhancement of the EC response in the relaxor ferroelectric system can be achieved by changing the chemical composition and approaching to the multiphase coexisting points at which the number of the coexisting phases led to maximize the entropy change, thus, inducing a giant EC response at relatively low electric field [59,60]. Moreover, BTSn ceramic presents the maximal EC response near the room temperature in a relatively wide range of temperatures, which makes it more attractive for practical cooling applications.

**Table 5.** Electrocaloric properties near the room temperature of BTSn and various lead-free ceramics.

| Ceramic | $T$ | $\Delta T_{max}$ | $\Delta E_{max}$ | $\zeta$ | Method | Ref. |
|---|---|---|---|---|---|---|
| | (°C) | (K) | (kV/cm) | (K.mm/kV) | | |
| BTSn11 | 52 | 0.71 | 25 | 0.284 | Indirect | This work |
| BTSn10 | 70 | 0.4 | 20 | 0.20 | Indirect | [59] |
| BTSn10.5 | 28 | 0.61 | 20 | 0.31 | Indirect | [58] |
| BTSn11 | 44 | 0.63 | 20 | 0.315 | Direct | [27] |
| BTSn12 | 52 | 0.45 | 20 | 0.225 | Indirect | [59] |
| BTSn14 | 18 | 0.38 | 20 | 0.19 | Direct | [27] |
| BTSn15 | 28 | 0.42 | 20 | 0.21 | Indirect | [59] |
| BZT-30BCT | 60 | 0.30 | 20 | 0.15 | Indirect | [61] |
| $Ba_{0.85}Ca_{0.15}Zr_{0.10}Ti_{0.90}O_3$ | 87 | 0.492 | 17 | 0.289 | Indirect | [29] |
| $Ba_{0.94}Ca_{0.06}Ti_{0.87}Sn_{0.125}O_3$ | 25 | 0.63 | 20 | 0.32 | Indirect | [62] |
| $Ba_{0.90}Ca_{0.10}Zr_{0.05}Ti_{0.95}O_3$ | 119 | 0.465 | 25 | 0.186 | Indirect | [30] |
| $Ba_{0.65}Sr_{0.35}TiO_3$ | 42 | 0.42 | 30 | 0.14 | Indirect | [63] |

### 3.5.2 Phenomenological theory using LGD

The temperature change, $\Delta T$, can also be calculated by using Landau-Ginzburg-Devonshire phenomenological theory, which can describe the macroscopic phenomena in ferroelectric materials in the vicinity of the phase transition [8]. For relaxor ferroelectrics, Pirc et al. [16] proposed a phenomenological theory model to calculate the electrocaloric effect. The Gibbs free energy can be stated as Eq. (10) in terms of electric polarization [16].

$$G = G_0 + \frac{1}{2}aP^2 + \frac{1}{4}bP^4 + \frac{1}{6}cP^6 + \cdots - EP, \quad (10)$$

Here, $a$, $b$, $c$, ..., etc. are the temperature-dependent phenomenological coefficients. In normal ferroelectrics, $a = \beta\ (T - T_0)$ where $\beta = (\varepsilon_0 \times C)^{-1}$ is a thermodynamic coefficient which is associated to the Curie constant $C$ and to the Curie-Weiss temperature ($T_0$). The entropy change is obtained from Eq. (11) [16]

$$\Delta S = -(\partial G / \partial T)_E = -\left(\frac{1}{2}\beta(T + \Delta T)P^2(E, T + \Delta T) - \frac{1}{2}\beta(T)\,P^2(0, T)\right). \quad (11)$$

The EC response can be expressed by using $\Delta T = -(T\Delta S/C_E)$ [16] as follows

$$\Delta T = \frac{T}{2C_E}[\beta(T + \Delta T)P^2(E, T + \Delta T) - \beta(T)P^2(0, T)], \quad (12)$$

where $C_E$ denotes the phonon part of the specific heat capacity of the material at given field $E$.

In principle, it is worth to note that Eq. (9) in Pirc paper is the self-consistent equation which should be solved numerically, since $\Delta T$ is coming in both sides of the equation [16]. Importantly, since $\Delta T << T$, Eq. (12) can be simplified to:

$$\Delta T = \frac{T}{2C_E}[\beta(T)P^2(E, T) - \beta(T)P^2(0, T)], \quad (13)$$

Fig. 6 compares the temperature-dependence of the estimated $\Delta T$ obtained via LGD theory with that deduced from the Maxwell relation under 25 kV/cm in BTSn ceramic. Both methods show the same trend with a maximum around FE-PE phase transition around 50 °C, then diminish continuously. The $\Delta T$ obtained by LGD was found to be 0.61 K, which is lower than that found by Maxwell relation (0.71 K). However, at the paraelectric phase transition, both methods yielded the same values. The apparent discrepancy of EC responses through both methods is related to the equation based on the LGD theory. The coefficient $a$ was considered to be linearly temperature-dependent for normal ferroelectrics whereas the coefficient $\beta$ is equal to $1/(\varepsilon_0 C)$. Note, however, that according to Pirc et al. [16], the coefficient $\beta$ is temperature-dependent in relaxor ferroelectrics, which could be the main reason for the difference between the results obtained by the indirect and the phenomenological approach [8,11].

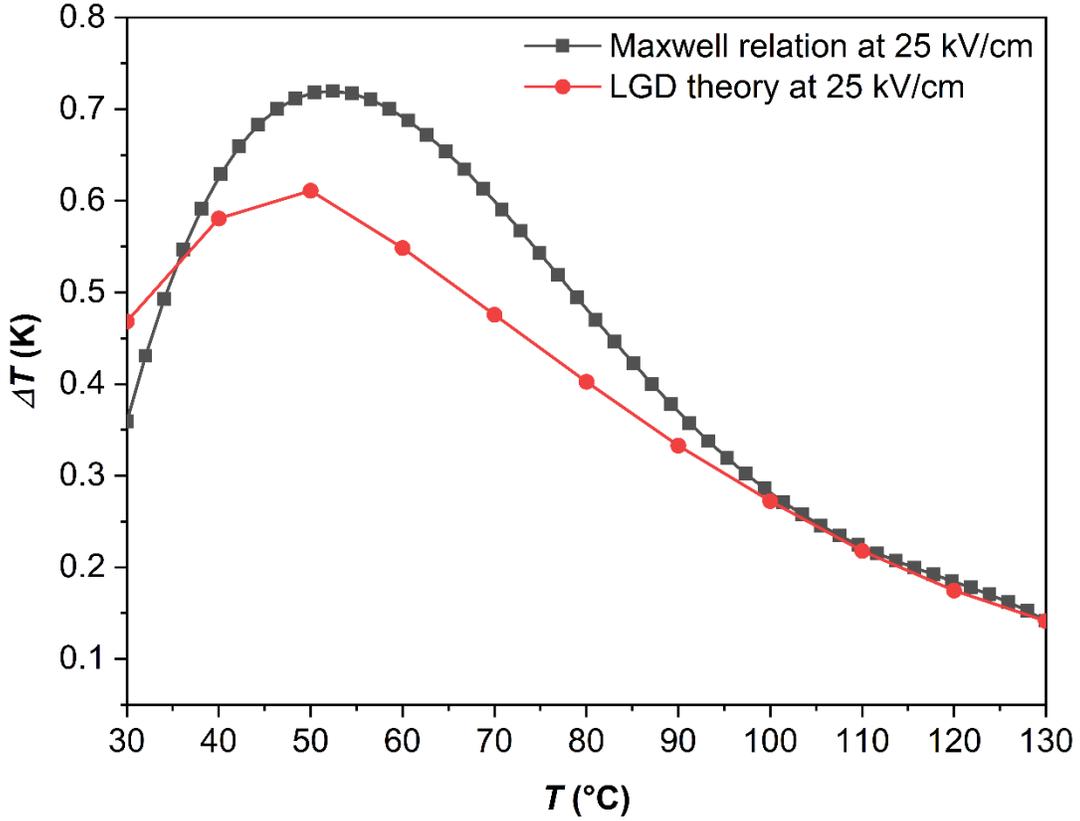

**Fig. 6.** Comparison of the thermal evolution of $\Delta T$ obtained by Maxwell relation and LGD theory under 25 kV/cm for BTSn ceramic.

## 4. Conclusions

The energy storage, electromechanical as well as electrocaloric properties in lead-free BaTi$_{0.89}$Sn$_{0.11}$O$_3$ (BTSn) ceramic under a low electric field of 25 kV/cm have been determined. Remarkably, at room temperature, enhanced dielectric ($\varepsilon_r$ = 11025), piezoelectric ($d^*_{33,ave}$ = 278 pm/V), and energy storage ($W_{tot}$ = 85.1 mJ/cm$^3$, $W_{rec}$ = 72.4 mJ/cm$^3$, $\eta$ = 85.07 %) properties were achieved. The temperature-dependence of the above-mentioned properties was also studied. For instance, large dielectric constant of 17390 was obtained at 41 °C, high recoverable energy storage ($W_{rec}$ = 84.4 mJ/cm$^3$) with an appreciable efficiency ($\eta$ = 91.04 %) were achieved at 60 °C. The electrocaloric effect was investigated by Maxwell relation and Landau-Ginzburg-Devonshire (LGD) phenomenological theory. The maximum EC response was found to be 0.71 K and 0.61 K, by Maxwell relation and LGD theory, respectively, around 50 °C. Hence, BTSn ceramic is a remarkable ferroelectric material for high-efficiency energy storage applications and also suitable for solid-state cooling devices.

## Acknowledgements

The authors gratefully acknowledge the generous financial support of CNRST Priority Program PPR 15/2015, the Slovenian research agency grants P1-0125, J1-9147, and the European Union Horizon 2020 Research and Innovation actions MSCA-RISE-ENGIMA and MSCA-RISE-MELON under the grant agreement No 778072 and 872631, respectively.